\documentclass[sigconf]{acmart}

\usepackage{booktabs}
\usepackage{multirow}
\usepackage{listings}
\usepackage{balance}

\AtBeginDocument{%
  \providecommand\BibTeX{{%
    \normalfont B\kern-0.5em{\scshape i\kern-0.25em b}\kern-0.8em\TeX}}}

\copyrightyear{2021}
\acmYear{2021}
\setcopyright{iw3c2w3}
\acmConference[WWW '21]{Proceedings of the Web Conference 2021}{April 19--23, 2021}{Ljubljana, Slovenia}
\acmBooktitle{Proceedings of the Web Conference 2021 (WWW '21), April 19--23, 2021, Ljubljana, Slovenia}

\acmPrice{}
\acmDOI{10.1145/3442442.3458605}

\acmISBN{978-1-4503-8312-7/21/04}

\begin{document}

\title{Demonstration of Faceted Search on Scholarly Knowledge Graphs}

\author{Golsa Heidari}
\affiliation{\institution{Leibniz University Hannover}\country{Germany}}
\email{golsa.heidari@stud.uni-hannover.de}
\orcid{0000-0002-5398-7086}

\author{Ahmad Ramadan}
\affiliation{\institution{Leibniz University Hannover}\country{Germany}}
\email{ramadan@stud.uni-hannover.de}
\orcid{0000-0002-3238-4315}

\author{Markus Stocker}
\affiliation{\institution{TIB Leibniz Information Centre for Science and Technology}\country{Germany}}
\email{markus.stocker@tib.eu}
\orcid{0000-0001-5492-3212}

\author{S\"oren Auer}
\affiliation{\institution{TIB Leibniz Information Centre for Science and Technology}\country{Germany}}
\email{auer@tib.eu}
\orcid{0000-0002-0698-2864}

\renewcommand{\shortauthors}{Heidari et al.}


\begin{abstract}
Scientists always look for the most accurate and relevant answer to their queries on the scholarly literature. Traditional scholarly search systems list documents instead of providing direct answers to the search queries. As data in knowledge graphs are not acquainted semantically, they are not machine-readable. Therefore, a search on scholarly knowledge graphs ends up in a full-text search, not a search in the content of scholarly literature. In this demo, we present a faceted search system that retrieves data from a scholarly knowledge graph, which can be compared and filtered to better satisfy user information needs. Our practice's novelty is that we use dynamic facets, which means facets are not fixed and will change according to the content of a comparison.
\end{abstract}


\maketitle

\section{Introduction}
Scholarly Knowledge graphs are knowledge bases for representing scholarly knowledge~\cite{Jaradeh2019}. Faceted search is a method that augments traditional search systems with a faceted exploration system, allowing users to narrow down search results by applying multiple filters based on the classification of the properties~\cite{Feddoul2019}. A faceted classification system lists each knowledge component along various dimensions, called facets, facilitating the classifications to be reached and managed in multiple forms.

Although faceted search is exceptionally beneficial for knowledge retrieval, search engines have not used it for the scholarly literature. Google Scholar is a well-known example: Despite its vast database, it has facets just on the date and, thus, limited support for refining queries. The next section briefly describes how implementing a faceted search over scholarly knowledge supports granular refinement of search queries.

\begin{figure*}[t]
  \includegraphics[width=\textwidth]{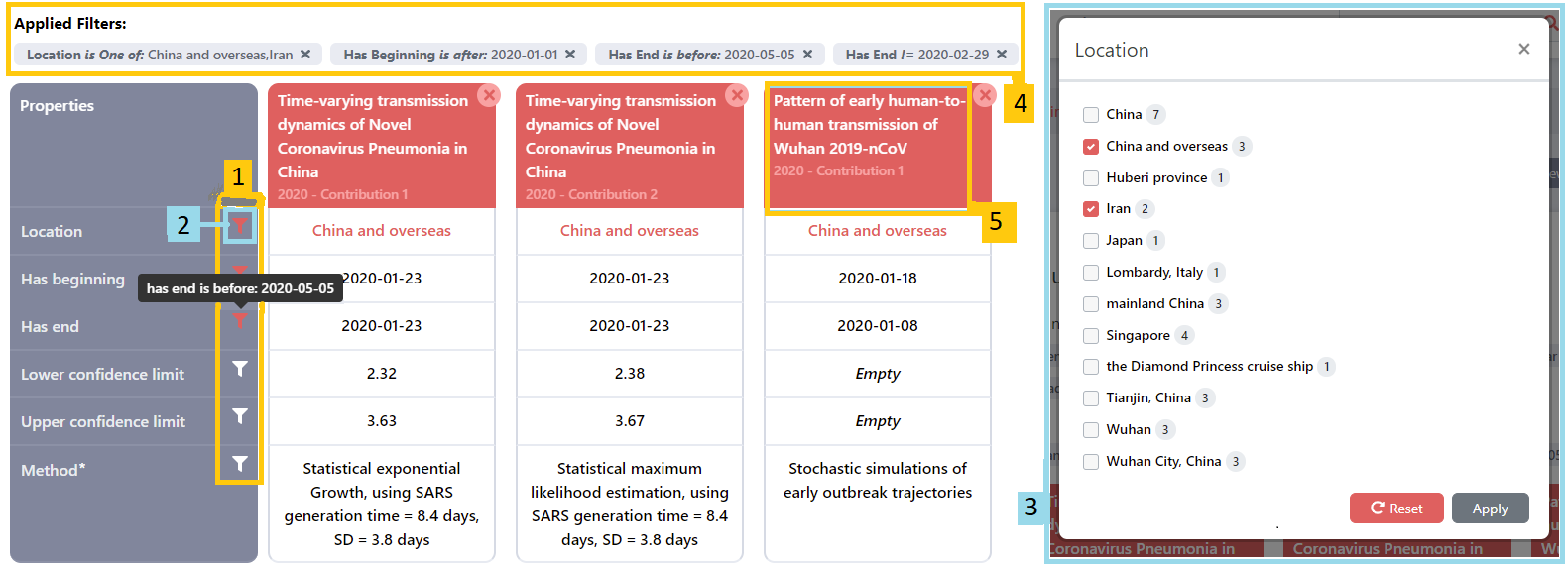}
  \caption{UI overview of faceted search use-case on COVID-19 contributions comparison in the ORKG. \\* 1) Filter icons to select the value for the properties. 2) If a filter icon is clicked, a dialogue box like (3) would appear. 3) A selection prompt of candidate facets. 4) Currently activated filters on the comparison. 5) Linked value objects (i.e., papers) that link back to all interconnected properties shown in the table. }
  \label{fig:ui}
\end{figure*}

\section{Approach}
We implemented our faceted search system on top of Open Research Knowledge Graph (ORKG) Comparisons. The ORKG\footnote{\url{https://gitlab.com/TIBHannover/orkg}} is an online service that represents research contributions (papers) in the form of an interconnected knowledge graph~\cite{Oelen2019Comparing} and enables the generation of tabular representations of contributions as comparisons, which is the main focus of our system.

The data about each paper is defined in properties that each one has a predefined template. These templates support the dynamic and automated construction of facets for ORKG comparisons. Facets work on different types of data throughout comparisons. For String properties (i.e., properties that have strings for values), a user can select one or more values among all. This is also supported by an auto-complete feature to help find candidate options. For numerical data, users may not only want to filter by a distinct value but also by a range. Hence, different operators can be selected for the filtering process, specifically greater or smaller than a specific amount. Furthermore, a user can exclude values. Similarly, operators can be applied for values of type date. In addition to including or excluding a date, a duration can be selected as a valid filtering criterion. A date picker is activated on date properties so a user can easily select the date on a calendar. For example, papers that used a special method of research and got special value(s) in a specific duration of a particular location and so on, can easily be discovered. 
Our system is empowered by these dynamic facets, which are inferred automatically from the property type, and the facets that would be different for each comparison, in contrast to other search systems which use just a predefined set of static facets. 

\autoref{fig:ui} depicts an example of the faceted search performed on a COVID-19 contribution comparison, which consists of 31 papers. When the filter icon is pressed, a dialogue box containing the relevant values for the property appears, thus enabling the user to choose some of the candidate values. When applying a filter, the colour of the filter icon is highlighted in red, and a tool-tip is displayed when hovering over the mentioned icon with the selected value(s). Additionally, all applied filters are indicated clearly on top of the table. The results are directly reflected on the screen.

Furthermore, the system provides the opportunity to save these configurations and the subset of retrieved data as a new comparison to the database, with a permanent URL that can be shared with other researchers and users. We provide a link to the system to enable independent testing and investigation.\footnote{ \url{https://www.orkg.org/orkg/comparisons}}

\section{Challenges}
What made the problem of faceted search challenging are the following points:
\begin{itemize}
    \item Knowledge graphs are heterogeneous by nature. Different KGs have different structure. Thus, they are not compatible with a strict search system. Various schemas and APIs make the exploration of federated systems even harder.
    \item Completeness matters. The more complete the database is, the more data would be discovered. Unfortunately, some well-structured systems suffer from a complete data source~\cite{heist2020knowledge}.
    \item Specifically concerning the ORKG, each paper is related to one or more research fields. Therefore, Finding the appropriate facet according to the user’s search expression is challenging.
\end{itemize}

The code of the system is publicly available and documented on GitLab.\footnote{ \url{https://gitlab.com/TIBHannover/orkg/orkg-frontend}}

\section{Conclusion}
Nowadays, knowledge graphs are central to the successful exploitation of knowledge available in the growing amount of digital data on the Web. Such technologies are essential to upgrade search systems from a keyword match to knowledge retrieval, which is vital for achieving the most relevant answer to a query, especially also in research. Despite remarkable gains by search portals, full-text scholarly search engines have remained ineffective. In this project, we implemented a faceted search system over a scholarly knowledge graph. To ease information retrieval, facets adapt to content. The more the knowledge graph is implemented in details, the search results would be more fine-grained. In future work, we will federate knowledge graphs to further improve dynamic faceted search. For instance, we intend to use GeoNames\footnote{\url{https://www.geonames.org/}} to enable spatial filtering on scholarly knowledge.

\begin{acks}
This work was funded by the Leibniz University of Hannover. The authors would like to thank Allard Oelen, Mohamad Yaser Jaradeh, and Kheir Eddine Farfar for helpful comments.
\end{acks}

\bibliographystyle{ACM-Reference-Format}

\bibliography{references}

\end{document}